\documentclass[12pt]{article}
\usepackage{graphicx}
\usepackage{color}
\newcommand{\beq}{\begin{equation}}
\newcommand{\eeq}{\end{equation}}
\newcommand{\bea}{\begin{eqnarray}}
\newcommand{\eea}{\end{eqnarray}}

\newcommand{\gsim}{\lower.7ex\hbox{$
\;\stackrel{\textstyle>}{\sim}\;$}}
\newcommand{\lsim}{\lower.7ex\hbox{$
\;\stackrel{\textstyle<}{\sim}\;$}}

\newcommand{\eod}{\end{document}}

\definecolor{verm}{rgb}{0.8,0.1,0.0}

\begin{document}
\thispagestyle{empty}
\vspace*{-22mm}

\begin{flushright}

UND-HEP-13-BIG\hspace*{.08em}10\\
Version 5.2 \\

\end{flushright}

\vspace*{1.3mm}

\begin{center}
{\Large {\bf Fundamental Dynamics \& Symmetries and for the future
-- like CP Violation \& EDMs}}

\vspace*{10mm}

{\bf I.I.~Bigi$^a$} \\
\vspace{7mm}
$^a$ {\sl Department of Physics, University of Notre Dame du Lac}\\
{\sl Notre Dame, IN 46556, USA}

\vspace*{-.8mm}

{\sl email address: ibigi@nd.edu}

\vspace*{10mm}

{\bf Abstract}\vspace*{-1.5mm}\\
\end{center}
Working with Kolya Uraltsev was a real `marvel' for me in general, but in 
particular about CP \& T violation, 
QCD \& its impact on transitions in heavy flavor hadrons and EDMs. The goal was -- 
and still is -- to define fundamental parameters for dynamics, how to measure them and compare SM forces with New Dynamics using the best tools including our brains. The correlations of them with accurate data  were crucial for Kolya. Here is a review of CP asymmetries in $B$, $D$ and $\tau$ decays, the impact of perturbative and non-perturbative QCD, about EDMs till 2013 -- and for the future.

\vspace{3mm}

\hrule

\tableofcontents
\vspace{5mm}

\hrule\vspace{5mm}

\section{About my collaboration with Kolya}
\label{MEMORY}

In 1988 Kolya, V.A. Khoze, A.I. Sanda and I wrote the article 
"The Question of CP Noninvariance -- as seen through the Eyes of Neutral Beauty" that was  published  in the World Scientific book "CP Violation" edited by C. Jarlskog \cite{1988BOOK}. 
I am very proud of that article;  
here are two of the best reasons for me: 
\begin{itemize}
\item 
After long discussions about this article with the Nobel prize winner Jack Steinberger 
at the ARCS2000 in the French alps, he smiled and said: `Very good work.'   
\item 
It discussed CP violation predicted with quarks and how they are 
affected by `hard' and `soft' re-scattering; I will come back to that.
\end{itemize}
Kolya and I had not met in person in 1988;  
that happened first in 1990 when he visited me at Notre Dame and we produced our second paper, 
namely "Induced Multi-Gluon Couplings and The Neutron Electric Dipole Moment" \cite{1990EDM}. A few months later when Kolya was back in the Leningrad Institute of 
Nuclear Physics, he sent me a Russian paper with a very similar title about this item  
and asked me, if I agree with the content\footnote{At that time I could read it in Russian; now you can read that article in English.}.
I did -- so  he put my name as an author and submitted it to Zh.Eksp.Teor.Fiz., where it was 
published \cite{SOV1991EDM}. I was very impressed 
by the work that he had done after the previous paper and I was honored (and still am) that 
Kolya put my name on it.  

Kolya and I spent long times together both at Notre Dame and CERN, in particular in the year of 1993/94, when we could work most of the night (and even in the week from Christmas to New Year, when the CERN rooms were cold) \footnote{Kolya's wife Lilya knew how to deal with Russian theorists; 
Kolya could not find a better wife, and he knew it. Kolya followed Lilya's expeditions to Kola Peninsula in the summers to help getting food by fishing.}. 
Vivek Sharma allowed us to use the rooms, computers and printers there 
\footnote{Vivek should get fair credit for that.}. It was a fruitful year. We had worked with Vainshtein and Shifman by email and phone about establishing 
Heavy Quark Symmetry (HQS) and its expansion (HQE) not only in principle, but also `practically'.  
Kolya had produced 
 new theoretical tools for non-perturbative QCD. He knew that hard theoretical work is needed to produce trustworthy predictions -- 
and that it often takes a lot of time. 
 He also said that in the end data are the supreme judges.  There is also a basic difference 
 between {\em pre}- and {\em post}-diction. One of the obvious examples is the history 
of the ratio of the lifetimes of beauty baryons vs. mesons. Kolya and his collaborators (and Voloshin as well)
stated many times that HQE gives a ratio of around 0.9 - 1.0 
 -- and 
were steadfast about their prediction when it was in obvious disagreement with the data 
during a decade or more. Kolya had thought a long time before arriving at 
\cite{LAMLIF}: 
\beq
\frac{\tau (\Lambda_b)}{\tau (B_d)} \simeq 1-\Delta_b \; , \;  \Delta _b \simeq 0.03 - 0.12 \; . 
\label{ratio}
\eeq
It is important to remember that the limit of HQE is unity here. Therefore our control (or lack) 
of nonperturbative QCD depends on $\Delta_b$, which shows large theoretical uncertainties\footnote{Voloshin preferred to say 
$\Delta _b \simeq 0.0 - 0.1$ \cite{VOLO}.}.
The most accurate data come from the LHCb collaboration \cite{LHCBRATIO}:
\beq
\frac{\tau (\Lambda_b)}{\tau (B_d)} = 0.974 \pm 0.006 \pm 0.004  \; . 
\eeq 
Both Kolya ({\em et al.}) \cite{LAMLIF} and Voloshin \cite{VOLO} were very happy about it -- but not surprised at all. 
This was a true {\em pre}diction which was very, very different from data for a long time. 
It is not obvious how much work and deep understanding of dynamics were needed to give these predictions -- but it is remarkable. 

Before I write about dynamics in some details, I want to say: Kolya was a 
wonderful person. I know he was very interested about art and history, as  obvious for a person who was born in Russia, worked in Italy, France, Japan and Siegen (the painter P.P. Rubens was born in Siegen, 
not in Holland!). 
Kolya with his family and I spent a day in the small city in Arezzo in Italy to see paintings, 
churches and architecture there. 
He liked long discussions about fundamental physics with passion and honesty. He was a true wonderful friend. He often asked me about my family's health situation (including my mother's one) and how he could help, when he was in a bad situation himself. I miss him so much. 

Here I give a review mostly of CP symmetry and its asymmetries, impact of QCD, some subtle points about EDMs and write about the future for the next ten years.

\section{CP symmetry and its violations}
\label{CPS}

In 1964 -- fifty years ago -- CP violation was found due to the existence of 
$K_L \to \pi^+\pi^-$. Okun explicitly listed the search for it as a priority for the future in his 1963 textbook \cite{OKUNBOOK} -- a true prophet, since he was the only one. It has been 
predicted in 1981 \cite{BS81} that sizable or even large CP violation should be found 
in $B^0 \to \psi K_S$ transitions when the CKM model of 
flavor dynamics is a least the leading source of it. The existence of the $B$ mesons had not been 
established then, never mind top quarks. After 1981 Sanda and I, and in parallel Uraltsev and collaborators,  worked out  the CP asymmetries in $B$ and $D$ mesons and  refined the theoretical 
tools, without communications between Russia on one side and West Europe/USA on the 
other side due to the `iron curtain'. It was predicted that CKM dynamics produce sizable or even large 
{\em indirect} CP violation in $B_d$ oscillation, but small one in the $B^0_s$ mixing. 

\subsection{Landscape of CP \& T violation between 1986 and 2013}

It was known that two neutral $B^0$ and $B^0_s$ mesons oscillate, but in different landscapes 
of $x=\Delta M_B/\Gamma_B$  due to SM dynamics. As pointed out by Azimov, Uraltsev \& Khoze \cite{AZIMOV}, indirect CP violation is small in $B^0_s$ oscillations.  
However  sizable 
CP asymmetries can occur in CKM suppressed $B^0_s$ decays, and therefore one has to look for them. They emphasized the 
hierarchy of neutral $B$ decays. Their quantitative predictions are based on the 
experimental claim that top quarks 
have been found with $M_t = 40 \pm 10$ GeV. It has been found that this claim was wrong, as history shows  
\footnote{When I found the statement from CERN outside our offices in the theoretical HEP group in Aachen, 
I looked at it and said: "They found it!" Peter Zerwas look at it, read it, thought for a few minutes and said: 
"It must be wrong, and I give you my reasons." Peter was correct as usual.}. Still Azimov, Uraltsev \& Khoze  had good reasons to be proud of this paper, since the basic idea is correct.

By 1987 the $B$ mesons had been found with 
$|V_{ub}| \ll |V_{cb}|$ and sizable $B_d$ oscillations \cite{ARGUS87}; 
indirectly they gave reasons for the existence of top quarks with  50 GeV $< m_t <$ 200 GeV. The long article \cite{1988BOOK} by Khoze, Sanda, Uraltsev and me discussed both indirect and direct CP violation in $B_d$, $B_s$ and $D^0$. The collaboration of these theorists happened by 
phones, emails and exchanging files between the US West Coast  and Russia. 

The 1988 article consisted of five Acts plus Prologue \& Epilogue:  
Act I. The Plots: CP Asymmetries in $B$ Decays; 
Act II. The Likely Hero: $B_d$ \& its Decays;  
Act III. The Dark Horse: $B_s$ \& its Decays; 
Act IV. The Dark Side -- Search Scenarios; 
Act V. Conclusions and Outlook.   

It focused on the following crucial points: 
(i) Three sides of  `the golden' CKM triangle are all of the order of $\lambda ^3$; it gives CP asymmetries between $\sim$10 \% and $\sim$80\%  for $B_d$ and $B^+$ transitions due to large angles \cite{MAILLOL}. 
(ii)  Another triangle allows to probe $B_s$ transitions. It gives one small angle, which  leads to about 5\% indirect CP violation  in $B_s$ oscillations due to the $\lambda ^2$ suppression, but it allows for large direct CP asymmetries.  
(iii) CP violation in $D$ decays are of the order of $\lambda ^4$ -- i.e. a 
few $\times 10^{-3}$ in the SM.  
(iv) There are `good' and `bad' signs; I will come back to that later.  
(v) We have to look for the impact of New Physics (NP) with higher accuracy and/or in rare decays. 
(vi) To have direct CP violation in two- and three-body final states (FS) one needs final state interactions (FSI), and 
strong forces provide them. 
(vii) Penguin diagrams \cite{PENG} can affect or even produce direct CP asymmetries in $B_{u,d,s}$ decays. 
However there are subtle points: penguin diagrams are formulated for quark states; 
to compare those predictions with measured data with hadrons one has to use the concept of `duality'.  

Penguin diagrams were introduced for kaon nonleptonic decays based on their connections with local operators \cite{SVZ}. In $B$ decays they affect inclusive final state 
with `hard' re-scattering. 
For CP asymmetry in exclusive decays one has to deal with `soft' re-scattering. Based on 
rough models 
we predicted $A_{\rm CP}(\bar B_d \to K^- \pi^+) \sim 0.1$.  It is a decent early prediction 
about subtle features of $B$ dynamics. Actually this article gave good predictions in general, namely: 
\begin{itemize}
\item 
$B_d$ transitions are the `hero' of true large CP violation in the SM as predicted in Act II. It was stated that in the future 
the angles $\phi_1$ in $B_d \to \psi K_S$ and $\phi_2$ in $B_d \to \pi^+\pi^-$ will be measured with sizable or even 
large values; the latter one will also show sizable direct CP asymmetry.  

CP violation in the $B$ system was established only in 2001 in $B_d \to \psi K_S$;  PDG 2013 gives:
\beq
S_{CP}(B_d \to \psi K_S) = {\rm sin} \phi_1 = + 0.676 \pm 0.021
\eeq 
Very recent Belle data also show \cite{BELLE2PI}: 
\bea
S_{CP}(B_d \to \pi^+\pi^-) = - 0.64 \pm 0.08_{stat.}\pm 0.03_{syst.}
\\
A_{CP}(B_d \to \pi^+\pi^-) = + 0.64 \pm 0.33_{stat.}\pm 0.03_{syst.}
\eea 
\item 
Indeed $B_s$ decays are `Dark Horse(s)' as in Act III: (i) The SM gives CP asymmetries in semi-leptonic decays significantly less than $10^{-4}$, while NP could produce it `closer' to 0.01. 
(ii) Indirect CP violation in 
$B_s \to \psi \phi$ could be seen around a few percent in the SM, while NP would reach the 
10 - 20 \% level as Sanda \& I had said before 2000. (iii) Direct CP asymmetries in CKM suppressed decays could be large.  

Very recent LHCb data confirm these 1988 predictions in subtle ways \cite{LHCBBS}; 
first:  
\beq
\frac{\Delta \Gamma (B_s)}{{\rm ps}} = 0.106 \pm 0.011 \pm 0.007    \; , \;   
\frac{\Gamma (B_s)}{{\rm ps}} = 0.661 \pm 0.004 \pm 0.006 \; , \; y_s \simeq 0.08
\eeq
Kolya and I were not sure (and I am still) about the small theoretical uncertainty about 
$\Delta \Gamma (B_s)$; Alex Lenz discusses it in his contribution. I focus on indirect 
CP violation: 
\beq
\phi_{\rm s}^{c \bar cs}|_{\rm data} = (0.01 \pm 0.07 \pm 0.01)\; {\rm rad}  \; \;
{\rm vs.} \; \; \phi_{\rm s}^{c \bar cs}|_{\rm SM} = (-0.0363 ^{+0.0016}_{-0.0015}) \; {\rm rad} 
\eeq
The data are close to the expected SM values -- but also consistent with sizable or even leading NP contributions. Furthermore I disagree with the 
uncertainty from the SM usually claimed in the literature; below I will explain why. 

\item 
Search scenarios for CP violation in neutral heavy mesons were discussed  in Act IV 
about the `Dark Side' of neutral beauty and charm mesons:
(i) It was emphasized that CP 
violation in inclusive decays is much smaller than  in exclusive ones. 
(ii)  CP asymmetries in $D^0$ decays were discussed. 
\item 
In Act V it was pointed out that, as of  1988, the connection between the observables and the underlying 
electroweak parameters was `obscured' by the impact of FSI. 
\item 
The epilogue gave important comments: if detailed data on $B$ decays could be compared 
with our rather accurate predictions, and failed them --   
there would be "no plausible deniability" that the CKM theory could no longer be maintained as the sole or even dominant source of CP violation -- therefore NP had to exist. 

Now the landscape has changed: the SM gives at least the leading source of 
CP violation in $B$ decays; therefore we have to go from `accuracy' to `precision'. 
That is the landscape upon us; it seems to me that `young' people working in HEP 
(both on the experimental and theoretical side) would like to produce such a transition (I hope). 
Furthermore, we need a deeper understanding of charm decays. 

\end{itemize}
We knew that the impact of penguin diagrams on hadronic FS 
and CP asymmetries is very subtle already in $B$ decays, and even  more in $D$ ones. 
It was discussed at a deeper level in 1990 by Dokshitzer and Uraltsev in Ref.\cite{DOK}, and at the DPF-92 meeting at Fermilab, where Kolya gave a talk on a short paper about FSI phases \cite{FSICP}. One needs only a short time to read it -- but a long time to think about the items like inclusive vs. exclusive transitions and `hard' vs. `soft' re-scattering.

From my own direct experience during this long period I know that Kolya was a real leader in probing CP violation in beauty and charm decays and in understanding 
the information given by data. 
Furthermore Kolya showed us that the first and second rounds are not enough -- one has to go further. 

\subsection{Duality -- the connection of quark and hadronic diagrams}
\label{DUAL}

The issue of `duality' between quarks and hadronic forces has been used in very different situations. Some are straightforward like for `jets', while 
others are subtle: flavor forces depend crucially on the non-perturbative aspects of QCD. 
Kolya and collaborators have worked on duality as a tool in HQE, mostly for beauty hadrons 
decays. 

It is not enough to give hand-waving statements there -- we can defend them with some 
accuracy to measure CKM angles and to compare inclusive vs. exclusive rates. It is not 
enough at all to compare sums of measured hadronic FS rates vs. parton model ones with quarks. One of the first papers to deal with this subtle issue dates back to 1986 \cite{VOLOSH}. 
It gives us insight into the inner structure of strong forces. We had discussed local 
vs. semi-local duality and how close one has to go to thresholds as discussed in Ref.\cite{HIGHPOWER}. 
It will be discussed elsewhere in this Memorial Book; still I list a few important references about 
Kolya's work about duality \cite{KOLYADUALITY}.  
For heavy quarks the ratios of lifetimes of baryons and mesons 
go to unity in the heavy quark limit like $\sim (\Lambda/m_Q)^2$ in HQE; 
thus the theoretical uncertainty is `sizable'.  The next steps 
are to measure $\tau (\Xi_b^0)$ and $\tau (\Xi_b^{-})$ with some accuracy. It was predicted \cite{VOLO}: 
$\tau (\Xi_b^0) \simeq \tau (\Lambda_b) < \tau (B_d) < \tau (\Xi_b^-)$. 
It shows how much we can control the impact of non-perturbative QCD in a semi-quantitative way on inclusive decays of beauty hadrons. 
Data told us we can reproduce the lifetimes of charm ones semi-quantitatively  -- is it just luck?

\subsubsection{Re-scattering \& CP violation \& CPT constraints}
\label{RESC}

It is important to think about the connections between {\em quark diagrams} and measured (or measurable) rates with {\em hadrons}. There are important, but subtle points: 
\begin{itemize}
\item 
In the quark world we use weak dynamics with $ b \bar q \to q_1\bar q_2 q_3 \bar q_4$ 
(even including `Weak Annihilation/Scattering'). Using the SM we deal with inclusive FS with $q_i = u,d,s$. Including QCD forces we use $m_u < m_d \ll m_s < \bar \Lambda$. The predictions 
are different due to iso-spin and $SU(3)_{fl}$ violations, however differences are small compared to 
$\bar \Lambda$ for inclusive rates. Measured inclusive FS consist of sums of hadrons; those show little effect of $SU(3)_{fl}$ violation.
\item 
The landscapes for {\em exclusive} non-leptonic decays are quite different. Two-, three- and four-body etc. 
FS can easily show sizable $SU(3)_{fl}$ violation and therefore about CP asymmetries. The impact are due to re-scattering with QCD dynamics, in particular soft re-scattering with non-perturbative forces. It can 
be calibrated by $M_K$ vs. $M_{\pi}$, the impact of chiral symmetry and its violations.

\end{itemize} 
The connection between the strength of re-scattering, CP asymmetries and CPT invariance has been discussed in Ref.\cite{1988BOOK}; it is explained in Sect. 4.10 in Ref.\cite{CPBOOK} with much more 
details, including CPT invariance following the history sketched above:
\bea
T(P \to a) &=& e^{i\delta_a} \left[ T_a + 
\sum_{a \neq a_j}T_{a_j}iT^{\rm resc}_{a_ja}  \right] 
\label{CPTAMP1} 
\\
T(\bar P \to \bar a) &=& e^{i\delta_a} \left[ T^*_a +
\sum_{a \neq a_j}T^*_{a_j}iT^{\rm resc}_{a_ja}  \right]   \; , 
\label{CPTAMP2}
\eea 
where amplitudes $T^{\rm resc}_{a_ja}$ describe FSI between $a$ and intermediate states $a_j$ 
that connect with this FS. 
Thus one gets for CP asymmetries: 
\beq
\Delta \gamma (a) = |T(\bar P \to \bar a)|^2 - |T(P \to a)|^2 = 4 \sum_{a \neq a_j} T^{\rm resc}_{a_ja} \, {\rm Im} T^*_a T_{a_j} 
\eeq
This CP asymmetry has to vanish upon summing over all such states $a$: 
\beq 
\label{cptinv}
\sum_{a} \Delta \gamma (a) =
4 \sum_{a}\sum_{a \neq a_j} T^{\rm resc}_{a_ja} {\rm Im} T_a^* T_{a_j} = 0 \; , 
\eeq
since $T^{\rm resc}_{a_ja}$ and Im$T_a^* T_{a_j}$ are symmetric and antisymmetric, 
respectively, in the indices $a$ and $a_j$. This shows that CPT invariance imposes equality also 
between {\em sub}classes of partial widths. There are important points: 
\begin{itemize}
\item
These equations show the non-trivial impact of re-scattering/FSI in general. 

\item
CP asymmetries in two-body FS give us `only' numbers. Those in three-body FS give us 
two-dimensional observables, namely measure Dalitz plots. 

\item 
In principle (with infinite data) one can probe `local' CP asymmetries. However we have to be realistic 
and use tools to reduce the numbers of observables. 
Therefore we probe `regional' asymmetries. 
We can use the definition of `fractional' asymmetry  
or `significance' one or others  \cite{MIRANDA1}. Furthermore it depends on the `landscape' of the FS 
where we choose it. It needs `judgement' based on 
our experience, namely on the impact of resonances and the differences 
between narrow and broad ones. It helps significantly to use chiral symmetry.

\end{itemize}
For this work I 
apply these amplitudes for FS with hadrons and resonances: 
$P \to h_1[h_2h_3] + h_2[h_1h_3] + h_3[h_1h_2] \Rightarrow h_1h_2h_3$.

The above equations apply to amplitudes of hadrons or quarks and also to boundstates of 
$\bar q_iq_j$ (or $q_iq_jq_k$). The crucial point is how to connect `measurable' hadronic amplitudes with 
quark and gluon ones. One can show that connection with diagrams; however in quantitative ways  it is subtle due to non-perturbative forces. To make it short: in the world of quarks our tools 
mostly focus on inclusive transitions, unless one can use other theoretical tools; it depends on our  
`judgement'. I will discuss it separately in $B$ and $D$ decays.

\subsubsection{Comments about `hard' vs. `soft' re-scattering}

There is a large difference between computing penguin diagrams and what they mean for beauty and charm decays (for different reasons) -- unlike for kaon transitions. In beauty decays one can 
calculate {\em inclusive} CKM suppressed FS with CP asymmetries based on `hard' re-scattering 
between FS with local operators. Penguin diagrams can give us the direction about correlations between hadrons in exclusive FS, but not in a truly quantitative way. In charm decays we have penguin diagrams about CP asymmetries, but less control over `soft' re-scattering as discussed in Ref.\cite{SOFTCHARM} 
in some details.

\subsection{CP violation via Higgs dynamics}
\label{HIGGSCP}

We have known for a long time 
that non-minimal Higgs models could not contribute sizably to $\epsilon$ and/or $\epsilon^{\prime}$ unless `our' world lives in very tiny corners of Higgs forces.  
In the 1997 book "Perspectives on Higgs Physics II"  
\cite{HIGGSBOOK} there is an article by Sanda, Uraltsev and me about "Addressing the Mysterious 
with the Obscure -- CP Violation via Higgs Dynamics". It focused on  
EDMs of neutron \& electrons \& atoms, T {\em odd} electron-nucleon interaction \& 
$K \to \mu \nu \pi$ 
and CP violation in $B$ \& $D$ \& top transitions. One neutral Higgs boson has been found in 2012, but no charged one (yet). 

 Non-minimal Higgs models are one of leading candidates for 
NP: they provide  us with a road to SUSY (and thus string theory) directly and indirectly. I will discuss it in the next subsection; however I will first comment about the measurable status of the real Higgs. 

The amplitude of the SM neutral Higgs state is 100\% scalar. ATLAS and CMS have established the existence of a neutral spin 0 boson $\Phi$ with a mass of 
$125.8 \pm 0.4(\rm stat.) \pm 0.4(\rm syst.)$  GeV in the FS of $2\gamma$, $l^+l^-$, 
$\bar bb$ quarks, $ZZ^*$ and $WW^*$ \cite{ATLASH}. 
We know that $\Phi$ is at least mostly a scalar boson, and pseudo-scalar contributions are at best subleading. 
Small pseudo-scalar amplitudes cannot produce sizable rates by themselves, but 
can sizably contribute to the interference with 
scalar SM ones -- i.e., the $j = 0$ $\Phi$ state can mix strongly  with the scalar amplitude  and weakly with the pseudo-scalar one to produce CP asymmetries \cite{BERN2}.
 Understanding the Higgs sector is an important project for very high luminosity runs at LHC and 
also for ILC. 

Another comment about production and decay of the established neutral Higgs boson: usually ATLAS and 
CMS base their analyses on the Higgs width predicted by the SM; however the impact of NP and 
Dark Matter can hide there. A new idea has appeared, namely to probe 
$pp \to H + X_1 \to `ZZ' + X_2 \to e^+e^- \mu^+\mu^- + X_3$ with 
$M(e^+e^- \mu^+\mu^-) > 130$ GeV that hardly depends on $\Gamma_H$; then one can 
compare the data from 126 GeV \cite{ELLISFERMI}.

\subsection{Probing CP asymmetries in the future}
\label{FUTCP}

We know that the SM gives at least the leading source of CP violation in $\Delta B \neq 0$ dynamics. Furthermore the neutral Higgs boson has been found now as expected with $M(H^0) \simeq 125$ GeV. 
On the other hand the usual reasons for NP exist,  namely:
\begin{itemize}
\item
We need NP to produce `us', namely huge matter vs. antimatter asymmetry now; 
forces based on the CKM matrix can{\em not} do it. 
\item
Theorists tell us extreme fine tuning is necessary to get 
electro-weak symmetry breaking below 1TeV, if NP is at a much higher scale,
like $10^{10}$ or $10^{15}$ or $10^{19}$ GeV. 
\end{itemize}
The data and our experimental colleagues tell us: 
\begin{itemize}
\item
The three neutrinos have different masses to give oscillations as measured.
\item
Working to get 
 `known' matter (and `us') is a sideshow in `our' universe, since it produces only  
around 4\% part of our universe. 
\item 
Dark matter gives around 23\% of it; there are several candidates like SUSY, but none 
is established (yet). 
\item
Dark energy gives around 73\% -- but what is it? 

\end{itemize}
Really a lot of work has  been done about `our' universe -- but we need even more. 
CP asymmetries can be based on the interference between SM amplitude and NP one. 
As usual, indirect searches for the impact of NP can reach much higher scales than direct ones due to inferences of two different  amplitudes.

NP can produce at best non-leading source of CP violation in $B$ mesons. Therefore one needs 
more data with better experimental and theoretical accuracies. First one focus on 
(quasi-)two-body FS. However I think we have to measure three- and four-body 
non-leptonic FS and their `topologies' like the two-dimensional Dalitz plots and the correlations 
between narrow and broad resonances. We need even more data and more tools, but the data give us much more information about the underlying forces. 
People who work in Hadro-Dynamics have produced and checked their technologies about 
$h_1 h_2 \to h_3 h_4$ scattering; now we can apply them to achieve a deeper understanding of 
fundamental physics. Also we know now that the usual Wolfenstein parameterization is very 
adequate for leading sources of CP violation, but not for non-leading ones. 

The situation is different for charm decays. The SM produces only small asymmetries in singly 
Cabibbo suppressed (SCS) decays and close to zero in doubly Cabibbo (DCS) ones. The limits from the data tell us that NP can produce only small asymmetries in SCS decays. For DCS 
decays we have little limits on CP asymmetries. However we need much larger productions 
of $D_{(s)}$ and $\Lambda_c$ states. Again we have to probe FS with three- and four-body FS; 
CPT invariance is usable there.

The SM cannot produce measurable CP asymmetry in $\tau$ decays beyond the well measured 
CP violation in $K^0 - \bar K^0$ oscillations. BaBar data show a CP asymmetry in 
$\tau^- \to \nu K_S\pi^- [+\pi^{0}]'s$ that is opposite to the predicted one -- but at only 2.9 sigmas: 
\bea
A_{\rm CP}(\tau ^+ \to \bar \nu K_S\pi^+)|_{\rm SM} &=& +(0.36 \pm 0.01) \% \; \; \; \cite{TAUBS} 
\label{TAU1}
\\
A_{\rm CP}(\tau ^+ \to \bar \nu K_S\pi^+ [+\pi^{0\; \prime} {\rm s}])|_{\rm BaBar2012} &=& -(0.36 \pm 0.23 \pm 0.11) \%
\; \; \cite{BABARTAU}
\label{TAU2}
\eea
Now available data probe only integrated CP asymmetries. It is important to probe 
regional CP asymmetries in $\tau ^- \to \nu [S=-1]$ FS; we have to wait for Belle II 
(and Super-Tau-Charm Factory if and when it will ever exist). As pointed out last year, it is important  to measure the correlations in $D^+ \to K^+\pi^+\pi^-/K^+K^+K^-$ etc. \cite{TAUD+}. 

Finally we have to probe correlations between {\em known} matter 
and candidates of {\em dark} matter in CP asymmetries and rare $B$ and $D$ decays -- if we get even more data with precison and understand underlying dynamics with better theoretical tools. 


\subsubsection{Better parameterization of the CKM matrix}

With three quark families one constructs six triangles with different shapes, but also 
the same area. Obviously one can construct them in several ways. One can do it by their three 
sides (or the ratios);  crucial contributions come from 
$|V_{cb}|$, $|V_{ub}|$, $|V_{td}|$, $|V_{ts}|$ etc. In particular vivid discussions are still happening on the comparison of $|V_{cb}|_{\rm excl}$ vs. $|V_{cb}|_{\rm incl}$ and 
$|V_{ub}|_{\rm excl}$ vs. $|V_{ub}|_{\rm incl}$.  
Kolya was a true leader in predicting these four classes of transitions and understanding the informations the data tell us about the underlying dynamics 
with accuracy; it is discussed in other articles in this Memorial Book. 

PDG and HFAG show the `exact' CKM matrix with three families of quarks. 
However experimenters and theorists do not use exact CKM matrix as you can see in their 
papers and talks. 

In Wolfenstein parameterization one gets six triangles that are combined into three classes with four parameters $\lambda$, $A$, $\bar \eta$ and $\bar \rho$ with 
$\lambda \simeq 0.223$. 
Those are probed and measured in $K$, $B$, $B_s$ and $D$ transitions: $A \sim 1$, but 
the two remaining ones are {\em not} of ${\cal O}(1)$: 
$\bar \eta \simeq 0.34$ and $\bar \rho \simeq 0.13$. It is assumed -- usually without mentioning -- that one applies them without expansion of $\bar \eta$ and $\bar \rho$. 
Obviously it is a `smart' parameterization with a clear hierarchy.  

Now we need a parameterization of the CKM matrix with more precision for non-leading sources 
in $B$ decays and very small one for CP asymmetries in $D$ decays with little `background' from SM. Several `technologies' were proposed,  like the one 
in Ref.\cite{AHN} with $\lambda$ as before, but $f\sim 0.75$, $\bar h\sim 1.35$ and 
$\delta _{\rm QM} \sim 90^o$. Now we get somewhat different six classes, and  it is more 
subtle for CP violation:   
\begin{eqnarray} 
\left(\footnotesize
\begin{array}{ccc}
 1 - \frac{\lambda ^2}{2} - \frac{\lambda ^4}{8} - \frac{\lambda ^6}{16}, & \lambda , & 
 \bar h\lambda ^4 e^{-i\delta_{\rm QM}} , \\
 &&\\
 - \lambda + \frac{\lambda ^5}{2} f^2,  & 
 1 - \frac{\lambda ^2}{2}- \frac{\lambda ^4}{8}(1+ 4f^2) 
 -f \bar h \lambda^5e^{i\delta_{\rm QM}}  &
   f \lambda ^2 +  \bar h\lambda ^3 e^{-i\delta_{\rm QM}}   \\
    & +\frac{\lambda^6}{16}(4f^2 - 4 \bar h^2 -1  ) ,& -  \frac{\lambda ^5}{2} \bar h e^{-i\delta_{\rm QM}}, \\
    &&\\
 f \lambda ^3 ,&  
 -f \lambda ^2 -  \bar h\lambda ^3 e^{i\delta_{\rm QM}}  & 
 1 - \frac{\lambda ^4}{2} f^2 -f \bar h\lambda ^5 e^{-i\delta_{\rm QM}}  \\
 & +  \frac{\lambda ^4}{2} f + \frac{\lambda ^6}{8} f  ,
  &  -  \frac{\lambda ^6}{2}\bar h^2  \\
\end{array}
\right)
+ {\cal O}(\lambda ^7)
\end{eqnarray}
\bea
{\rm Class\; I.1:}&&V_{ud}V^*_{us} \; \; \;  [{\cal O}(\lambda )] + V_{cd}V^*_{cs} \;  \; \;  [{\cal O}(\lambda )] + 
 V_{td}V^*_{ts} \; \; \; [{\cal O}(\lambda ^{5\& 6} )] = 0   \\ 
{\rm Class\; I.2:}&& V^*_{ud}V_{cd} \; \; \;  [{\cal O}(\lambda )] + V^*_{us}V_{cs} \; \; \;  [{\cal O}(\lambda )] + 
V^*_{ub}V^*_{cb} \; \; \; [{\cal O}(\lambda ^{6 \& 7} )] = 0    \\
{\rm Class\; II.1:}&& V_{us}V^*_{ub} \; \; \;  [{\cal O}(\lambda ^5)] + V_{cs}V^*_{cb} \;  \; \;  [{\cal O}(\lambda ^{2 \& 3} )] + 
V_{ts}V^*_{tb} \; \; \; [{\cal O}(\lambda ^2  )] = 0   \\ 
{\rm Class\; II.2:}&& V^*_{cd}V_{td} \; \; \;  [{\cal O}(\lambda ^4 )] + V^*_{cs}V_{ts} \; \; \;  [{\cal O}(\lambda ^{2\& 3})] + 
V^*_{cb}V^*_{tb} \; \; \; [{\cal O}(\lambda ^{2 \& 3} )] = 0  \\
{\rm Class\; III.1:}&& V_{ud}V^*_{ub} \; \; \;  [{\cal O}(\lambda ^4)] + V_{cd}V^*_{cb} \;  \; \;  [{\cal O}(\lambda ^{3\& 4} )] + 
V_{td}V^*_{tb} \; \; \; [{\cal O}(\lambda ^3  )] = 0   \\ 
{\rm Class\; III.2:}&& V^*_{ud}V_{td} \; \; \;  [{\cal O}(\lambda ^3 )] + V^*_{us}V_{ts} \; \; \;  [{\cal O}(\lambda ^{3\& 4})] + 
V^*_{ub}V^*_{tb} \; \; \; [{\cal O}(\lambda ^4 )] = 0
\eea 
One finds the same pattern as from Wolfenstein parametrization, namely `large' CP asymmetries in Class III.1, 
sizable ones in Class II.1 and `small' one in Class I.1. However, the pattern is not so obvious, and it is similar only in a semi-quantitive way:  

(i)  In Class III.1 triangle one usually calls the two angles $\phi_1/\beta$  \& $\phi_3/\gamma$.
They are measured in CP asymmetries in $B_d \to \psi K_S$ \& $B^+ \to D_+K^+$ 
decays due to interference between two contributions one gets from CKM dynamics. Adapting the refined parametrization 
one finds that CKM dynamics produce $S(B_d \to \psi K_S) \sim 0.72$ as largest possible value for CP asymmetry with $\delta _{\rm QM} \simeq 100^o - 120^o$ to compare with the measured
\beq 
S(B_d \to \psi K_S) \sim 0.676 \pm 0.021 \; .
\eeq
When correlations with $\phi_2/\alpha$ and $\phi_3/\gamma$ point to 
$\phi_1/\beta \simeq 75^o - 90 ^o$ one gets 
$S(B_d \to \psi K_S)= \sin2\phi_1 \simeq 0.62 - 0.68$!
Therefore it seems that CKM dynamics give very close to `maximal' SM CP violation. However the situation is more subtle -- as discussed next.  

(ii)
We are searching for non-leading sources of CP violation in $B$ transitions. 
NP's impact could `hide' there in "SM predicted" CP asymmetries. 
`Data' given by HFAG, for example, are averaged over values of $|V_{ub}/V_{cb}|$ from 
inclusive and exclusive semileptonic $B$ decays;
actually the `central' value is closer to $|V_{ub}|_{\rm excl}$ rather than the larger 
$|V_{ub}|_{\rm incl}$. It is quite possible that the theoretical uncertainties about extracting  
$|V_{cb}|$, $|V_{ub}|$ and 
$|V_{ub}/V_{cb}|$ from $B \to l \nu \pi$ vs. $B \to l \nu D^*$ are much larger than claimed; 
some details are told about it in Refs.\cite{SCH}. A new idea using 
dispersion relations and chiral symmetry (in a smart way) came up very recently, namely to extract $|V(ub)|$ from data on $B \to l \nu \pi^+\pi^-$ \cite{KANG}; it probes the impact of broad scalar resonances. It gives us more roads to understand the underlying dynamics. One can think also about measuring 
$B_s \to l^+\nu K_S\pi^-$ and $B\to l \nu K \bar K$ and how much you can use chiral symmetry.

(iii) The information on NP from  data now and in the future  has to be based on 
accuracy and the correlations among different FS in several $B$, $D$ and $K$ transitions and rare 
decays.  In particular we can probe Class I.1 with the  tiny rates of  $K \to \pi \nu \bar \nu$ rates. The theoretical uncertainties are under control in that case; the hope is to produce enough data.


(iv) We have to probe correlations with different FS based on CPT invariance.  
The best fitting of the data do not give us the best information about the underlying dynamics.

\subsection{`Catholic' road to NP -- three-body final states}
\label{3BODY}

For $D/B \to P_1P_2P_3$ or $\tau \to \nu P_1P_2$ decays there is a single path to  `heaven', 
namely asymmetries in the Dalitz plots. One can rely on relative rather than 
absolute CP violation; thus it is much less 
dependent on production asymmetries. However one needs a lot of statistics -- and 
robust pattern recognition. 

\subsubsection{CP asymmetries in $B^{\pm}$ decays}

Data of CKM suppressed $B^+$ decays to charged three-body FS show not surprising rates
\bea
{\rm BR}(B^+ \to K^+\pi^-\pi^+) &=&(5.10 \pm 0.29 ) \cdot 10^{-5}  \\  
{\rm BR}(B^+ \to K^+K^-K^+) &=&(3.37 \pm 0.22 ) \cdot 10^{-5} \; .
\label{SUPP}
\eea
LHCb data show sizable CP asymmetries {\em averaged} over the FS 
with correlations \cite{LHCb028}: 
\bea
\Delta A_{CP}(B^{\pm} \to K^{\pm} \pi^+\pi^-) &=&  
+0.032 \pm 0.008_{\rm stat} \pm 0.004_{\rm syst}
[\pm 0.007_{\psi K^{\pm}}]  
\nonumber \\
\Delta A_{CP}(B^{\pm} \to K^{\pm} K^+K^-) &=&   
- 0.043 \pm 0.009_{\rm stat} \pm 0.003_{\rm syst}
[\pm 0.007_{\psi K^{\pm}}] \; .
\label{AVER1}
\eea 
It is not surprising that these CP asymmetries come with {\em opposite} signs, 
due to the CPT invariance constraint in eq.(\ref{cptinv}).
 Furthermore there are also large {\it regional} CP asymmetries which refer to a particular region of the phase space: 
\bea 
A_{CP}(B^{\pm} \to K^{\pm} \pi^+\pi^-)|_{\rm regional} &=& + 0.678 \pm 0.078_{\rm stat} 
\pm 0.032_{\rm syst}
[\pm 0.007_{\psi K^{\pm}}] 
\nonumber
\\
A_{CP}(B^{\pm} \to K^{\pm} K^+K^-)|_{\rm regional} &=& - 0.226 \pm 0.020_{\rm stat} \pm 0.004_{\rm syst}
[\pm 0.007_{\psi K^{\pm}}] \; .
\label{SUPP3}
\eea 
The `regional' CP asymmetries in the LHCb data mean here: 
(i)  
positive asymmetry at low $m_{\pi ^+\pi ^-}$ below $m_{\rho^0}$; 
(ii)  
negative asymmetry both at low and high $m_{K^+K^-}$ values.  These statements make very good 
sense. However I want to emphasize we need some thinking and judgment about 
the definitions of regional asymmetries and to go beyond analyses of the best fitted data.  
One has to remember that scalar resonances (like $f_0(500)/\sigma$ \& $\kappa$) produce broad ones that are not described by Breit-Wigner parametrization; instead they can be described by dispersion relations (or other ways). At the qualitative level one should not be surprised. Probing the topologies of Dalitz plots with accuracy one might find the existence of NP.  Most of the data come along the frontiers, while the centers are practically empty. 
Therefore interferences happen on few places, and regional asymmetries are much larger than averaged ones -- but so much? We have to remember that the final goal is to find non-leading sources of CPV 
in $\Delta B \neq 1,2$. 

One can look at even more CKM suppressed three-body FS: 
\bea
{\rm BR}(B^+ \to \pi^+\pi^-\pi^+) &=&(1.52 \pm 0.14 ) \cdot 10^{-5}
\\ 
{\rm BR}(B^+ \to \pi^+K^-K^+) &=&(0.52 \pm 0.07 ) \cdot 10^{-5} \; .
\label{MORESUPP}
\eea
One might guess that penguin diagrams have a smaller impact on $b\to d$ than on $b\to s$ 
comparing Eqs.(\ref{MORESUPP}) with Eqs.(\ref{SUPP}). 
On the other hand one also expects smaller impact on three- or more body FS than in two-body ones due to 
chiral symmetry. 

LHCb has shown these averaged and `regional' CP asymmetries \cite{PRL112}:
\bea
A_{CP}(B^{\pm} \to \pi^{\pm} \pi^+\pi^-) &=& + 0.117 \pm 0.021_{\rm stat} 
\pm 0.009_{\rm syst}
[\pm 0.007_{\psi K^{\pm}}] 
\nonumber \\
A_{CP}(B^{\pm} \to \pi^{\pm} K^+K^-) &=& 
- 0.141 \pm 0.040_{\rm stat} \pm 0.018_{\rm syst}
[\pm 0.007_{\psi K^{\pm}}] 
\label{AVER2}
\\
\Delta A_{CP}(B^{\pm} \to \pi^{\pm} \pi^+\pi^-)|_{\rm regional} &=&  
+0.584 \pm 0.082_{\rm stat} \pm 0.027_{\rm syst}
[\pm 0.007_{\psi K^{\pm}}]  
\label{SUPP7} 
\nonumber \\
\Delta A_{CP}(B^{\pm} \to \pi^{\pm} K^+K^-)|_{\rm regional} &=&   
- 0.648 \pm 0.070_{\rm stat} \pm 0.013_{\rm syst}
[\pm 0.007_{\psi K^{\pm}}] \; .
\label{SUPP2}
\eea
Again it is not surprizing that these asymmetries come with opposite signs. 
However there are two very interesting statements about the data shown \cite{PRL112}: 
\begin{itemize}
\item
$B^{\pm} \to \pi^{\pm}\pi^-\pi^+$ decays show CP asymmetries both positive signs with $m^2_{\pi^+\pi^-}> 15 \, {\rm GeV}^2$ and 
$m^2_{\pi^+\pi^-} < 0.4 \, {\rm GeV}^2$.

\item
On the other hand we find negative CP asymmetry in 
$m^2_{K^+K^-} < 1.5 \, {\rm GeV}^2$.

\end{itemize}
We need more data  --  they will appear `soon' --, find other regional asymmetries and work on correlations with other FS. Importantly we need more thinking to understand what the data tell us about the underlying dynamics 
including the impact of non-perturbative QCD. It seems that the landscape is even more 
complex than said before and show the impact of really broad resonances. 

There will be `active' discussions about the impact of CPT invariance, namely the duality -- 
averaged and regional transitions -- between the worlds of hadrons and quarks. There are several reasons to expect that the impact of penguin diagrams is large.
However there is a quantitative question now: 
How can CP asymmetries in Eq.(\ref{AVER2}) be larger by a factor of three than in 
Eq.(\ref{AVER1}) etc.? Also there are subtle questions, namely the definition of `regional' transitions: 
The best-fit result does not give us the best information about underlying dynamics, in particular 
about non-leading sources; we have to think deeper and use several theoretical tools.  
At first we need some good judgment to define `regional' asymmetry with finite data. 
Later we can test our judgment with even more -- but still finite -- data and correlations with 
other data. Still we need thinking -- model independent analyses are not always an 
excellent idea.

We have to think about the impact of penguins diagrams on {\em exclusive} rates and 
CP asymmetries. 
It shows the impact of penguins/re-scattering diagrams, since the FS with $\Delta S \neq 0$ 
are larger than with $\Delta S = 0$. However one can remember that penguins {\em operators} show only 
hard re-scattering and focus on inclusive decays.   First one probes 
averaged CP asymmetries, but later regional ones in the Dalitz plots and probe the correlations with different FS as shown above. Such procedures have been suggested and simulated in the case of three-body FS in $B^{\pm}$ decays \cite{MIRANDA1} as second step -- but this is not the final step in my view. The 
meaning of the analysis `being model independent techniques' is very complex with finite data with 
non-perturbative QCD about non-leading source of CP violation. One needs other theoretical tools like chiral symmetry and/or dispersion relations based on data with low energy collisions and/or correlations with other 
transitions.

\subsubsection{CP asymmetries in $D^{\pm}_{(s)}$ (\& $\tau^{\pm}$) decays}

So far no CP violation has been established in charm hadron decays. CPV can well be probed 
with two-body FS, but also in three-(and four-)body ones with more data like 
$D \to K_S \pi \pi$. A visionary paper by Azimov \& Iogansen about direct CPV in two-body FS
was published in 1981  \cite{IOG}. 

Probing three-body SCS and DCS gives more information about fundamental forces. 
It was pointed in 1989 \cite{GOLD} in general. One can disagree on several details, however 
it is important to think about our tools. 
$D^{\pm}$ has two all charged three-body FS on the SCS level -- namely 
$D^{\pm} \to \pi^{\pm} \pi^+\pi^-/\pi^{\pm} K^+K^-$ \cite{MIRANDA2} -- 
and also on the DCS one -- $D^{\pm} \to K^{\pm} \pi^+\pi^-/ K^{\pm} K^+K^-$. 
$D_s^{\pm}$ has two ones on the SCS level -- $D^{\pm}_s \to K^{\pm} \pi^+\pi^-/ K^{\pm} K^+K^-$ -- however only one for DCS level -- $D^{\pm}_s \to K^{\pm}K^{\pm}\pi^{\mp}$.

As stated above, for SCS FS the SM gives small `background' for CPV and very close to zero about DCS ones. We have to probe FS with broad resonances -- in particular scalar ones like 
$f_0(500)/\sigma$ and $\kappa$ -- and their interferences. The `landscapes' of Dalitz plots in charm decays are different from $B$ decays, because the central regions of phase space are not empty.

There may be a sign -- maybe -- of NP in $\tau$ decays about 
{\em averaged} CP asymmetries, see Eqs. (\ref{TAU1},\ref{TAU2}). 
It is crucial to probe regional ones. Furthermore one has to measure correlations with $D^{\pm}_{(s)}$ decays \cite{TAUD+}.

\subsection{`Protestant' road to NP -- four-body final states}
\label{4BODY}

There are several ways to probe CPV in four-body FS and to differentiate the impact of SM vs.\ NP: the landscapes are even more complex, while our theoretical toolbox is 
smaller so far. On the hand, when we will have more data on charm and beauty decays, it 
will enhance -- I hope -- the interests of young theorists (maybe from hadrodynamics) 
to  produce new tools to probe four-body FS. I focus on charm decays. One can compare 
T {\em odd} moments or correlations in $D$ vs. $\bar D$.  For example one has to measure the angle $\phi$ between the 
planes of $\pi^+-\pi^-$ and $K - \bar K$ and describe its dependence 
\cite{CPBOOK, TAUD+}: 
\bea
\frac{d\Gamma}{d\phi} (D \to K \bar K \pi^+\pi^-) &=& \Gamma_1 {\cos^2}\phi + \Gamma_2 {\sin^2}\phi +\Gamma_3 {\cos}\phi\, {\sin}\phi
\\ 
\frac{d\Gamma}{d\phi} (\bar D \to K \bar K \pi^+\pi^-) &=& \bar \Gamma_1 {\rm cos^2}\phi + \bar \Gamma_2 {\rm sin^2}\phi -\bar \Gamma_3 {\rm cos}\phi \,{\rm sin}\phi
\eea
The partial width for $D[\bar D] \to K \bar K \pi^+\pi^-$ is given by $\Gamma _{1,2} [\bar \Gamma _{1,2}]$: $\Gamma_1 \neq \bar \Gamma_1$ 
and/or  $\Gamma_2 \neq \bar \Gamma_2$ represents direct CPV in the partial width. 
$\Gamma_3$ and $\bar \Gamma_3$ represent T {\em odd} correlations; 
by themselves they do not necessarily indicate CPV, since they can be induced by strong FSI; however \cite{RIOMANI,CPBOOK}:
\beq 
\Gamma_3 \neq \bar \Gamma_3  \; \; \to \; \; {\rm CPV} 
\eeq   
Integrated rates give $\Gamma_1+\Gamma_2$ vs. $\bar \Gamma_1 + \bar \Gamma_2$; 
the integrated {\em forward-backward} asymmetry 
\beq
\langle A\rangle = 
\frac{\Gamma_3 - \bar \Gamma_3}{\pi (\Gamma_1+\Gamma_2+\bar \Gamma_1+\bar \Gamma_2)}
\eeq
gives full information about CPV. One could disentangle $\Gamma_1$ vs. $\bar \Gamma_1$ and 
$\Gamma_2$ vs. $\bar \Gamma_2$ by tracking the distribution in $\phi$. If there is a {\em production}  
asymmetry, it gives global $\Gamma_1 = c \bar \Gamma_1$, $\Gamma _s = c \bar \Gamma _2$ 
and $\Gamma_3 = - c \bar \Gamma_3$ with {\em global} $c \neq 1$. Furthermore one can applying these observables to $D[\bar D] \to 4 \pi$ (with CPT invariance) 
and later for $D^+ \to K^+\pi^-\pi^+\pi^-$ vs. $D^- \to K^-\pi^+\pi^-\pi^+$. There are different 
definitions of the angle between the planes of two hadrons. 

Of course, there are other `roads' to probe CP asymmetries in four-body FS with one-dimensional 
observables and compare them using correlations. 
We have learnt from  the history of 
$K_L \to \pi^+\pi^- \gamma ^*  \to \pi^+\pi^- e^+e^-$, 
where Seghal \cite{SEGHAL1,SEGHAL2} really predicted CPV there around 14\% based on 
$\epsilon_K \simeq 0.002$, where 
leptons have spin. It helps to discuss that situation in more details with unit vectors: 
\beq
\vec n _{\pi} = \frac{\vec p_+ \times \vec p_-}{|\vec p_+ \times \vec p_-|} \; , \;
\vec n _{l} = \frac{\vec k_+ \times \vec k_-}{|\vec k_+ \times \vec k_-|}  \; , \;
\vec z = \frac{\vec p_+ + \vec p_-}{|\vec p_+ + \vec p_-|} \\
\eeq 
\bea
{\rm sin} \phi = ( \vec n _{\pi}  \times \vec n _{l} ) \cdot \vec z \; [CP=-,T=-] &,& 
{\rm cos} \phi = \vec n _{\pi}  \cdot \vec n _{l} \; [CP=+,T=+]\\
\frac{d\Gamma}{d\phi} &\sim & 1 -(Z _3\,  {\rm cos} 2\phi  + Z_1\, {\rm sin}2\phi)
\eea
Then one measures asymmetry in the moments:
\beq
{\cal A}_{\phi} = 
\frac{(\int_0^{\pi/2} - \int_{\pi/2}^{\pi}+\int_{\pi}^{3\pi/2}-\int_{3\pi/2}^{2\pi})\frac{d\Gamma}{\phi}}
{(\int_0^{\pi/2} + \int_{\pi/2}^{\pi}+\int_{\pi}^{3\pi/2}+\int_{3\pi/2}^{2\pi}) \frac{d\Gamma}{\phi}}
\eeq
There is an obvious reason to probe the angle between the 
$\pi^+\pi^-$ and $e^+e^-$ planes based on $K_L \to \pi^+\pi^- \gamma ^*$ 
or $K^0 \to \pi^+\pi^- \gamma ^*$ vs. $\bar K^0 \to \pi^+ \pi^- \gamma^*$. 

However the situation for non-leptonic $D$ decays is more complex for several reasons; 
therefore one can use:
\bea
\frac{d}{d\phi} \Gamma (H_Q \to h_1h_2h_3h_4) &=& |C_Q|^2 - 
[B_Q\,  {\rm cos} 2\phi  + A_Q\, {\rm sin}2\phi] = \\
&=&|C_Q|^2 - [B_Q\,  (2 {\rm cos}^2 \phi  -1) + 2A_Q\, {\rm sin}\phi \, {\rm cos}\phi]  \\
\frac{d}{d\phi} \Gamma (\bar H_Q \to \bar h_1\bar h_2\bar h_3\bar h_4) &=& 
|\bar C_Q|^2 - [\bar B_Q\,  {\rm cos} 2\phi  - \bar A_Q\, {\rm sin}2\phi ]= \\  
&=&|\bar C_Q|^2 - [\bar B_Q\,  (2 {\rm cos}^2 \phi  -1) - 2\bar A_Q\, {\rm sin}\phi \, {\rm cos}\phi]
\eea
Obviously the landscapes are more `complex' 
\beq
\Gamma (H_Q \to h_1h_2h_3h_4) = |C_Q|^2 \; \; \;  {\rm vs.} \; \; \; 
\Gamma (\bar H_Q \to \bar h_1\bar h_2 \bar h_3\bar h_4) = |\bar C_Q|^2  
\eeq
For these moments one gets: 
\beq
\langle A_{\rm CPV}^Q \rangle = \frac{2(A_Q - \bar A_Q)}{|C_Q|^2+|\bar C_Q|^2}  \; ; 
\eeq
i.e., no impact from the $B_Q$ and $\bar B_Q$ terms. 
Furthermore one wants to probe semi-regional asymmetries like:
\beq
A_{\rm CPV}^Q|_a^b =
\frac{\int _a^b d\phi \frac{d\Gamma}{d\phi}- \int_a^b d\phi\frac{d\bar \Gamma}{d\phi}}
{\int _a^b d\phi\frac{d\Gamma}{d\phi}+ \int_a^b d\phi\frac{d\bar \Gamma}{d\phi}}
\eeq
where $B_Q$ and $\bar B_Q$ contribute. Again, the main point is not to choose which gives the 
best fitting one as discussed about $K_L \to \pi^+\pi^- e^+e^-$ \cite{SEGHAL1,SEGHAL2}.

I want to emphasize that many-body FS give us more information about the underlying dynamics. However we have not yet the best tools to get it quantitatively. More data 
will attract theorists to think about it.  

\subsection{Dealing with final states interactions}
\label{NABIS}

Tools about FSI in three-body FS have been produced in the last 15 years, for instance dispersion relations 
 based on low energy collisions with strong forces \cite{DISPREL}. 
Chiral symmetry is a good tool for probing FS for pions. However their impact 
and the connections of CPT are subtle for $\pi K \Leftrightarrow \pi K$ and 
$K\bar K \Leftrightarrow K \bar K$. For four-body FS we need more thinking -- but it is very 
important both on the theoretical and experimental side.

\section{Intermezzo: QCD \& the strong CP problem}
\label{NPQCD}

Very shortly I comment about the extraction of $V_{cb}$ and $V_{ub}$ and their correlation with CP asymmetries as discussed above and with EDM below. 

Comparing $|V_{cb}|$ and $|V_{ub}|$ from inclusive and exclusive semi-leptonic $B$ decays is a 
very `hot' item.  
It is crucial to constrain the `golden' CKM triangle (and the `second triangle' for $B_s$ transitions) 
with accuracy or even precision. As Kolya stated in his last conference talk in  November 2012 (using refined theoretical technologies like BPS and {\em non}-local correlations), he found no different values for $|V_{cb}|_{\rm incl}$ and $|V_{cb}|_{\rm excl}$; however it does not mean that the angles might not show NP in 
CP asymmetries. 

There is no local `competitor' with QCD to describe strong forces. 
However, the landscape of QCD forces is more complex and its connection with global symmetries and their violations. Usually we use QCD as a tool to find CP and T violation 
in $B$, $K$, $D$, top quarks, $\tau$ and neutron and leptons decays due to weak or superweak forces. 
Of course, there are very good reasons to probe the features of the strong forces in details.

It was pointed out in 1976 by 't Hooft \cite{HOOFT} that the dimension four operator 
$G \cdot \tilde G$ -- with $G$ gluon field strength tensor -- can be added to the QCD lagrangian. If one `decides' that the coefficent for this operator is zero, then quantum corrections will come back with non-zero value $\bar \theta$ . Thus strong forces violate both P and T invariance. 
Similarly, chiral invariance for massless quarks is no longer 
conserved in quantum field theory. 
Strong CP \footnote{Old problems (like old soldiers) never die -- they just fade away!} and chiral problems are furthermore intertwined 
by including also weak dynamics. The neutron EDM is described by an operator 
with dimension five. Thus its dimensionful coefficient $d_N$ can be calculated as a finite quantity, 
in particular for 
$d_N \sim {\cal O}[(e/M_N)(m_q/M_N)\bar \theta ] \sim {\cal O} (10^{-16} \bar \theta )$ e-cm 
\cite{BALUNI}. Data give limits about $d_N$ leading to $\bar \theta < 10^{-9}$ or less -- 
an `un-natural' limit as seen by most in high-energy physics. The Peccei-Quinn symmetry can make it `natural' \cite{AXION}. 
No axion has been found, and many members of our 
community thought that the `dawn' of the axions go the their `dusk'. However it seems to me that 
members of the cosmology community see a much more important role of axions, but not in the old version:  Peccei-Quinn symmetry might be broken also by UV dynamics; 
once axions were produced in the early universe, they constitute (part of) the DM in the present 
universe \cite{KIWOON}.  Renaissance  for (refined) axions?

\section{Subtle working for EDMs} 
\label{EDM}

So far CP and T violations have been established in $\Delta S\, \& \, \Delta B \neq 0$ transitions, 
but not in flavor diagonal ones (except `our' existence). There are excellent reasons 
to probe EDM deeper and deeper in many different states: from elementary leptons 
($e$, $\mu$, $\tau$), to very complex states (heavy atoms and nuclei), with neutron, proton and deuteron in between. It tells us that the {\em ratio} of NP vs. SM effects can be huge.  
However one goes after tiny effects in subtle environments. It needs long time commitments 
of the experimental groups (and the funding agencies). If an EDM has been 
found and established, it would be a wonderful achievement. Then we have to understand the 
features of the underlying NP. It would be a golden mine for theorists. 
We have not found them yet. However theorists might help experimenters  to continue their hard work with good ideas to find other systems with non-zero EDMs and later about correlations 
with other ones. 
 
Again Kolya's broad horizon is apparent: 
he worked on EDMs and Higgs dynamics in the 1980's, then on CP asymmetries in heavy quark transitions, next on the impact 
of perturbative and non-perturbative QCD and then on EDMs again. I know he had thought many times, as you had seen during and afterwards discussions even for talks given by other 
people; you knew whether Kolya was attending a seminar or not -- it was obvious.

\subsection{Early era}

Kolya had worked about neutron EDM with A.A. Anselm in 1984 \cite{ANS}. 
I had  my first meeting with Anselm at SLAC around 1986; he came to my office after a talk 
I had given about CP violation in $B_d$ decays and just mentioned a paper about neutron EDM due to 
Higgs exchanges. He told me politely that most people neglected quark interaction with 
neutral Higgs bosons. It was claimed that these contributions should be proportional to the third power of light {\em current} quarks. However the nucleon coupling with a neutral Higgs boson depends on nucleon mass at low momenta and does not vanish in 
the chiral limit. Higgs exchange could not give sizable contribution to CP violation in kaons transitions, but still could produce neutron EDM around $10^{-22}$ e-cm -- i.e., that prediction exceeds the experimental limit by at least around two orders magnitude. There were very subtle statements to understand that; I had never heard that before. Therefore I read that paper \cite{ANS} right away and was very impressed by it. Later I met Anselm at least twice at Winter  Schools close to Moscow and always enjoyed talking and discussing with him. Kolya was a graduate student of Anselm and co-autor of the 1984 paper.   
 
Kolya and I had produced the first paper in `person' when Kolya was invited to the physics department at Notre Dame in 1990 with the title "Induced Multi-Gluon Couplings and 
the Neutron EDM" \cite{1990EDM}. Let us assume that in the future this or another idea will be established `natural' to make $\bar \theta < 10^{-9}$ or less. 
Then we can deal with a 
challenge that on the surface is hardly connected with the $G \cdot \tilde G$ problem. 
Non-minimal Higgs models can produce neutron EDM close to the experimental limit while 
contributing very little to $K_L \to \pi \pi$ due to the emergence of the 
$G^2 \tilde G$ operator. In 1990 Kolya generalized and refined these arguments: 

\noindent 
(a) the operator $G^2 \tilde G$  is induced in different classes of models for 
CP violation. It had also been noted by several authors that typically `sizable' effects arise there \cite{WEINBERG}. The CKM dynamics produces a coefficient of the $G^2 \tilde G$ operator 
that is utterly tiny. 

\noindent
(b) We had discussed a new method for estimating the relevant matrix element, namely 
$\langle N|i\bar q \sigma_{\mu \nu}q F_{\mu \nu} |N\rangle$ as induced by $G^2 \tilde G$: 
it yields a result that is considerably smaller than Weinberg's estimate. 

\noindent
(c) We refined the findings from other authors that QCD radiative corrections suppress rather than enhance of the impact the operator $G^2 \tilde G$. 

We discussed the one-, two- and three-loop situations including QCD radiative corrections. 
It was non-trival work to get three conclusions: 
(i) We found strong evidence that contribution from $G^2 \tilde G$ operator is quite unlikely 
to be cancelled by additional Peccei-Quinn term. 
(ii) Finding neutron EDM larger than $10^{-31}$ e-cm is a clear sign of the existence of NP 
in T violation -- but not of its features.  
(iii) There are even more (theoretical and experimental) reasons to probe EDM 
for electrons, atoms, nuclei, $\mu$, $\tau$ etc. coming forward. 

An excellent 1991 article, "The electric dipole moment of the electron" by 
Bernreuther \& Suzuki \cite{BERN}, focused on the electron EDM and how to probe it in atoms and molecules; 
it also discussed the connection with neutron EDM including Kolya's work. 

\subsection{Around 2000 - 2013}

The prospects of probing EDMs as a direct sign of T violation became even more exciting around 2000 with new ideas and more tools and technologies \cite{RITZREV} 
(and also for sociology reasons, since these experimental collaborations are relatively  small).  
One could see that in conferences and workshops -- in particular in the 
"Flavor in the Era of the LHC"  CERN Workshop, November 2005 - March 2007; 
it produced a long and very good proceedings  published in Eur.Phys.J. C (2008) 
with three sections \cite{CERNFLAVOUR}. I have enjoyed and learnt from them, in particular EDM and g-2 miniworkshop on Oct. 9 - 11, 2006 \cite{CERNEDM}. Many discussions happened in `public' or `private' -- and Kolya liked that also.  
We need more data, more technologies -- and more thinking for leptons, quarks and gluon dynamics.

No EDM has been found yet anywhere in this (relatively) huge landscape in different `dimensions', namely to probe EDMs in neutron, protons, nucleis, atoms, molecules, charged leptons etc. \cite{JAPANREV}. 
The ACME collaboration 
has given limit on electron EDM: $|d_e| < 8.7 \cdot 10^{-29}\; e \cdot$ cm \cite{ACME}. I find it 
exciting to read how experimental physicists did it. Furthermore it attracts more theorists to 
think about probing EDMs in different directions (and some of them come back as before  
including connections about axions) \cite{SONNY,DEKENS}. We need new ideas as before.

As said before Kolya was a true leader in discussions about the importance of probing EDMs more and more in very different situations. Kolya had produced two papers together with 
Th. Mannel published in 2012/13. In Ref.\cite{2012MANNEL} they pointed out that 
the neutron EDM can be generated in the SM already by tree diagrams due to boundstate effects 
without short-distance penguin diagrams. It produces non-zero chiral limit and does not depend 
on the difference $m_s - m_d$; they estimated around $d_N \sim 10^{-31}$ e-cm -- 
a value similar to hand-waving arguments given in \cite{1990EDM}.  
In the longer paper \cite{2013MANNEL} they gave more details how they came to two statements: 
(i) The landscape of neutron EDM is described not by effective CP-odd operators of lowest dimension, but by non-trivial interplay of different amplitudes at low energy scales like 
1 GeV for two $\Delta C=1$ \& $\Delta S=0$ four-quark operators. (ii) Those operators can  
be probed in $D$ decays. 

Kolya and Thomas and many physicists (like me) were excited to find 
CPV in charm transitions for the first time in $D^0 \to K^+K^-$ vs. $D^0 \to \pi^+\pi^-$; 
furthermore those data seem for many theorists to be beyond what the SM can generate.   
However more LHCb data did not confirm this CPV in charm decays. 
Even so I think in the future CPV will be established in $\Delta C \neq 0$ transitions. My main point is: even non-established data can lead us to think deeper about fundamental forces. These two papers are good examples:  
{\em tree} diagrams can produce TV/CPV in boundstates of quarks and come back to an old, but still not mature question, namely the impact of penguin diagrams \cite{PENG} in general and in particular about nucleon EDMs how they depend on long-distance strong forces. 

Again they dealt with subtle points that most people prefer to forget: 

\noindent 
(i) We have to `understand' dynamics with measurable parameters, 
in particular `complex' impact on baryons transitions. 

\noindent
(ii)  We have to talk about the connection with `constituent' vs. `current' quarks 
as discussed before \cite{ANS}, but with more tools. 

\noindent 
(iii) We have to understand why several chiral suppression of light quarks can be vitiated in 
composite hadrons like nucleons. 

\noindent 
(iv) It is possible that the impact of `heavy' quarks loops is sizable or even important for nucleon EDMs and for boundstate effects. 

\noindent
(v) It is even more subtle to discuss the connection between neutron EDM and direct CP asymmetries in 
$B$, $D$ and $\mu$ and $\tau$ decays -- and their connections with Dark Matter. 

\noindent 
(vi) It is important to think about boundstate effects without short-distance dynamics from penguins. 

I had worked on that item in the past 
-- and have learnt 
so much from Kolya about fundamental dynamics. 

\subsection{Future era}

As mentioned before here (and  many other places like in Ref.\cite{CERNEDM}) we have to find non-zero EDMs in leptons, neutron, deuteron, atoms, molecules etc. for  several reasons: 
\begin{itemize}
\item 
The SM produces tiny `backgrounds' in all these situations. 
\item 
EDMs affect many landscapes and with different correlations.

\item 
It is a wonderful challenge for experimenters to apply their tools in a new situation or produce a new technology. It helps `inventiveness'. 
Young experimenters will enjoy much more working in small groups than in 
huge collaborations. 

\end{itemize}
Neither charged Higgs or a second neutral Higgs have been found, and the known one  is at least 
mostly a scalar. 
Fans of SUSY like me do not give up that this theory exists in our world -- but not   
in the mass region which LHC can probe directly. 

Obviously SUSY cannot solve all problems for fundamental forces together. 
However I do not think that our world prefers the minimal version of SUSY. 
Non-minimal versions can produce EDMs that can be measured in the future. Probing EDMs 
can be competitive with the reach of $\epsilon_K$ \cite{POPRECENT}. 

In the future experiments at FNAL will measure $(g - 2)_{\mu}$ with more data (and I hope 
also the  muon EDM later) and at the J-PARC Hadron Experimental Facility (Japan) the combined  measurements of $(g - 2)_{\mu}$ and $d_{\mu}$ using very new tools. Even if the second 
experiment fails to reach its goals, the community would learn so much and 
will come up with new ideas and new technology that can be applied in the `real' world. 
That way some of us respond to well known Beckett's skepticism\footnote{As said by Samuel Beckett: "Ever tried. Ever failed. No matter. Try again. Failed again. 
Fail better."}.
In one way one can compare 
$d_e < 105 \times 10^{-29}$ e-cm (from PDG) or the recent value 
$d_e < 8.7 \times 10^{-29}$ e-cm \cite{ACME} with 
$\delta [F_2(0)/2m_e] \sim 2\times 10^{-22}$ e-cm derived from 
$\delta [(g-2)/2] \sim 10 ^{-11}$ or $d_{\mu} = (-0.1 \pm 0.9) \times 10^{-19}$ e-cm. 

I have been thinking and working about EDMs and leptonic dynamics as signs of the features of NP and correlations with CP asymmetries in particular in charm hadrons and top quarks. The last two published 2012/2013 papers from Kolya gave me and others new `directions'. 
Three very recent papers \cite{POPRECENT,BERN2,POSP} point out that partners of SUSY 
might `easily' exist at mass scales of ten of TeV and above. The best way to probe those mass scales is to measure EDMs. I am not saying that LHC cannot find NP -- but we have to think about it. 

\section{Kolya's impact in the past, present and for the future}
\label{SUM}

Kolya has worked for around 35 years about fundamental dynamics in many landscapes  
and showed his broad horizon: impact of Higgs state(s), CP and T violation, perturbative 
and non-perturbative QCD and the correlations 
between landscapes (in particular the subtle ones). He always showed that first and second wins are  
not enough -- one has to get deeper. 
He always liked the connection with experimental colleagues, explaining why theoretical predictions are good or bad and where more theoretical work is needed. He also showed that real theorists do not act as `slaves' of the data of the time. Sometimes predictions are correct based on good theoretical tools (and a lot of thinking and working), while data are different; however eventually the data move closer and closer to good real {\em pre}dictions -- like the ratio of $\tau (\Lambda_b)/\tau (B^0)$ as mentioned above. 

\begin{figure}[h!]
\begin{center}
\includegraphics[width=6cm]{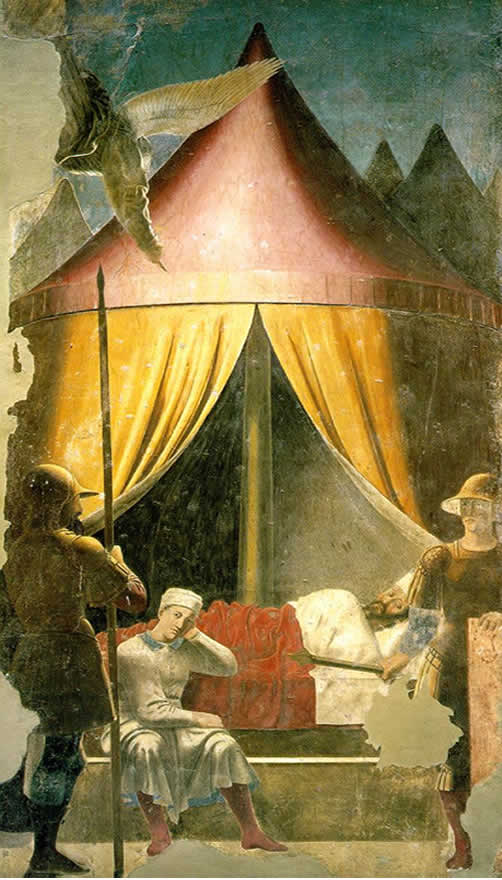} 
\end{center}
\caption{Dreaming in different dimensions} 
\label{fig:AREZZO}
\end{figure}

I give another `personal' comment about Kolya. During a workshop in Florence Kolya, Lilia, Gennady \&  
I went to Arezzo, a small city southeast of Florence to see architectures and churches; 
we saw paintings by Piero della Francesa produced around 1455 A.D. -- in particular about the dream of 
Constantine (the Great) Fig.\ref{fig:AREZZO}. 
Usually one is allowed to see it at most for half on hour, but due to Kolya's tenacity \& Gena's ability to  speak in good Italian language we saw \& discussed it close to two hours. I will never forget that experience.

\vspace{10mm}

{\bf Acknowledgments:} This work was supported by the NSF under the grant number 
PHY-1215979.



\begin{thebibliography}{99}
\bibitem{MAILLOL}
If a reader  is interested, like Kolya, in art and its connection with symmetry, she/he can see 
`Musee d'Orsay: Aristide Maillol, The Mediterranean/Thought' about "interlocking triangles" 
and "return to order".
\end{thebibliography}
\end{document}